\documentclass[%
 reprint,
nofootinbib,
nobibnotes,
bibnotes,graphicx,graphics,
 amsmath,amssymb,
 aps,stmaryrd,
]{revtex4-1}
\usepackage{natbib} 
\usepackage{xtab,afterpage}

\usepackage{soul}
\usepackage{tikz}

\newcommand{\comment}[1]{}
\usepackage{tikz}
  \newlength\squareheight
\usepackage{wasysym} 
\usepackage{graphicx}
\usepackage{dcolumn}
\usepackage{bm}

\begin{document}

\preprint{Draft}

\title{An exact formula for percolation on higher-order cycles}

\author{Peter Mann}
\email{pm78@st-andrews.ac.uk}
\author{V. Anne Smith}%
\author{John B.O. Mitchell}
\author{Christopher Jefferson}
\author{Simon Dobson}
\affiliation{School of Computer Science, University of St Andrews, St Andrews, Fife KY16 9SX, United Kingdom }
\affiliation{School of Chemistry, University of St Andrews, St Andrews, Fife KY16 9ST, United Kingdom }
\affiliation{School of Biology, University of St Andrews, St Andrews, Fife KY16 9TH, United Kingdom }

\date{\today}

\begin{abstract}
We present exact solutions for the size of the giant connected component (GCC) of graphs composed of higher-order homogeneous cycles, including weak cycles and cliques, following bond percolation. We use our theoretical result to find the location of the percolation threshold of the model, providing analytical solutions where possible. We expect the results derived here to be useful to a wide variety of applications including graph theory, epidemiology, percolation and lattice gas models as well as fragmentation theory. We also examine the Erd\H{o}s-Gallai theorem as a necessary condition on the graphicality of configuration model networks comprising higher-order clique sub-graphs.
\end{abstract}

\pacs{Valid PACS appear here}
\maketitle


\section{Introduction}
\label{sec:introduction}
Bond percolation on graphs is a process in which edges are randomly removed with some probability, $T$. As $T$ is reduced to some critical value, $T^*$, the graph exhibits a second-order phase transition and fails to be globally connected. The size of the GCC, as well as the location of the critical point, are important quantities within the percolation process. Percolation has not only inherent theoretical interest but is also important for various applications across many disciplines \cite{newman_2019,cohen_havlin_2010,desesquelles_2011,davila_escudero_lopez_dorso_2007,PhysRevC.76.054603,TRAUTMANN2005407,kuzmin_pleshakov_2019,cimino_cychosz_thommes_neimark_2013,PhysRevA.30.520}. Perhaps the most prominent utilisation is the study of diseases spreading through structured populations with transmission probability $T$. In this instance, the GCC is isomorphic to the outbreak size of the disease while the critical bond occupation probability is the epidemic threshold. It is well understood how to extract the properties of graphs using the generating function formulation \cite{newman2001rga,Newman2002SpreadOE,molloy_reed_1995,newman_2019,cohen_havlin_2010}. In its original form, it is assumed that there are no closed-loops or cycles among the edges of the graph; it is locally tree-like everywhere. When this condition is true, or approximately true, the generating function formulation yields excellent results compared to simulation. However, if the network fails to be locally tree-like, then the formulation must be modified to describe correctly the emergent properties of percolation. Newman \cite{PhysRevE.68.026121} provides an early analytical breakthrough in the study of graphs with closed loops. Within the generating function formulation, the next theoretical milestone is by Miller and Newman in 2009 who independently studied 3-cliques along with tree-like edges \cite{PhysRevE.80.020901,PhysRevLett.103.058701}. Shortly thereafter Karrer and Newman \cite{karrer_newman_2010} developed a general framework that addressed the study of larger subgraphs; however, it was determined that a crucial quantity, which we denote by $g_\tau$, could only be determined by an exponentially slow exhaustive enumeration of states. This quantity is the probability that a node remains unattached to the GCC despite its involvement in a cycle of topology $\tau$. Allard \textit{et al} \cite{Allard_2012,allard_hebert-dufresne_young_dube_2015} developed a comprehensive and versatile technique based on recursive formulas to determine the percolation properties numerically through iteration. Within the spirit of these developments, Mann \textit{et al} developed an analytical approach that approximates the $g_\tau$ expression to high accuracy \cite{2020arXiv200606744M,mann2020random} affording an equation-based treatment of percolation on arbitrary subgraphs. It remains that the percolation properties can be found exactly, but slowly through Karrer and Newman's method, exactly but recursively though Allard \textit{et al}'s method, or approximately but analytically though Mann \textit{et al}'s method. 

In this paper, we develop exact analytical expressions for homogeneous subgraphs; that is, cycles whose nodes are all degree-equivalent to one another. We present these equations for simple cycles and cliques; however, we hope it is clear how the method can be extended to other homogeneous classes that arise between these limiting examples. Application of our counting method to inhomogeneous cycles (cycles that contain nodes with different degrees) can readily be performed; however, the final expression depends on the details of the subgraph. The method is most similar to \cite{PhysRevE.68.026121,PhysRevE.80.020901,PhysRevLett.103.058701,karrer_newman_2010} and the polynomials we develop herein appear very similar to those found by \cite{PhysRevE.68.026121}, although we provide closed form expressions. 

\section{Background}
\label{sec:background}

It is necessary to review both the generating function formulation and the configuration model in order to progress with contents of this paper \cite{newman_2019,cohen_havlin_2010}. The framework is based on the degree distribution, $p(k)$, which is the probability of choosing a node of degree $k$ from the graph. Two generating functions are introduced that generate (i) the probability of choosing a node at random from the network
\begin{equation}
    G_0(z) = \sum_{k=0}^\infty p(k)z^k
\end{equation}
and (ii) the distribution of degrees of a node reached by following a randomly chosen edge 
\begin{equation}
    G_1(z) = \frac{G_0'(z)}{G_0'(1)}
\end{equation}
Defining $u$ as the probability that a neighbour is unattached to the GCC, the probability that a node fails to become attached through a single edge is $g_2 = 1-T+uT$, which is the sum of the probability that the edge was not occupied, $1-T$, and the probability that it \textit{was} occupied, but the neighbour was unattached to the GCC, $uT$. The quantity $u$ can be found as the solution to a self-consistent expression \cite{Newman2002SpreadOE}
\begin{equation}
    u=G_1(g_2)\label{eq:g1}
\end{equation}
The expected size of the GCC, $\mathcal S$, is then given by $\mathcal S= 1-G_0(g_2)$. The critical point can then be found by perturbing around $u=1$ which corresponds to $\mathcal S=0$ since $G_0(1)=1$. Expanding Eq \ref{eq:g1} with a Taylor series we have $u = 1 + uTG_1'(1) + \mathcal O(u^2)$, from which we find $T^*=1/G_1'(1)$ \cite{Newman2002SpreadOE,molloy_reed_1995,newman_2019}.

The configuration model is a method that can be used to create a particular random graph from an ensemble of degree equivalent, uncorrelated random graphs. In the model, the nodes of the graph are assigned an integer, drawn at random from the degree distribution, which indicate its degree. The degree sequence $\{k\}=k_1,k_2,\dots,k_{\mathcal N}$, where $\sum_i k_i=2E$ for a network of $\mathcal N\in \mathbb Z$ nodes and $E\in \mathbb Z$ edges, is a sequence of the degrees of the nodes and is typically displayed in descending order such that $k_1\geq k_2 \geq \dots \geq k_{\mathcal N}$. However, not all degree sequences are valid, or \textit{graphic}, such that some sequences of integers cannot be used to create a graph. The Erd\H{o}s-Gallai theorem (EGT) states that in addition to the handshaking lemma (HL), $\sum_i k_i=2E$, a sequence is graphic if and only if the Erd\H{o}s-Gallai inequality (EGI) 
\begin{equation}
    \sum_{i=1}^{n} k_i \leq n(n-1)+\sum_{i=n+1}^{\mathcal N} {\min}(k_i,n)\label{eq:EGT}
\end{equation}
holds for $n\in [1,\mathcal N-1]$. It is trivial to construct degree sequences that satisfy the HL (EGI) but do not satisfy the EGI (HL) and are thus not graphic. For instance, with $\mathcal N=3$ and $\{k\}=\{(1),(1),(1)\}$ the inequality in Eq \ref{eq:EGT} is satisfied but the sum of degrees is not even whilst $\{k\}=\{(2),(0),(0)\}$ satisfies the lemma but not Eq \ref{eq:EGT}.

To construct the networks, node $i$ is inserted $k_i$ times into a list for all $i\in \mathcal N$ which is then shuffled. Pairs of nodes are then drawn at random and connected together. In the limit of large and sparse networks, the probability that the construction process chooses pairs that are either already connected through another edge or belong to the same node is vanishingly small. The networks generated according to this process are locally tree-like and contain no short-range loops; they are also absent of degree-correlations.

\section{Graphicality of joint degree sequences}
\label{sec:graphicality}

The original configuration model described in section \ref{sec:background} was extended by Newman to incorporate triangular clustering \cite{PhysRevE.68.026121,PhysRevE.80.020901,PhysRevLett.103.058701} and subsequently higher-order subgraph motifs \cite{karrer_newman_2010}. In this model the degree distribution is replaced by a joint degree distribution that describes a node's involvement in higher-order cycles such as triangles, squares, 4-cliques etc. For instance, a node that is involved in $s$ ordinary edges and $t$ triangles is specified by joint degree $(s,t)$ and the usual degree is recovered from $k=s+2t$. Similarly, the joint degree of a node that is a part of $s$ ordinary edges, $t$ triangles, $v$ squares and $w$ 4-cliques is given by $(s,t,v,w)$ and occurs with probability $p(s,t,v,w)$, its ordinary degree is recovered from $k=s+2t+2v+3w$, a Diophantine condition \cite{10.1093/comnet/cnw011}. In the extended configuration model it is important to note that the cycles are independent of one another, in much the same way that simple edges are in the original model. This means that the accidental formation of a 4-clique during triangle construction through the choosing of two nodes that are already involved in a triangle vanishes with large and sparse networks. Thus, upon considering the characteristic size of each motif, the extended configuration model regenerates the locally tree-like property of the subgraphs. The probability of edge sharing between independent cycles is dependent on the number of nodes and triangles in the cycles for a given number of cycles, however. 

The degree sequence of a configuration model network is a sequence of tuples
\begin{equation}
    (s_1,t_1,\dots,\tau_1), \dots ,(s_{\mathcal N},t_{\mathcal N},\dots,\tau_{\mathcal N}),
\end{equation}
and as with ordinary edges, not all sequences lead to the successful creation of networks and we now consider necessary conditions on a joint degree sequence in order that it is graphic. It is natural to  separate and order the joint sequence as $s_1\geq s_2 \geq \dots \geq s_{\mathcal N}$ for the ordinary edges, $t_1\geq t_2 \geq \dots \geq t_{\mathcal N}$ for the triangles (and so on). 
It is clear that the EGT (the EGI and the HL) must still hold among the overall degrees of the model for the joint degree sequence to be graphic. However, the EGT is no longer sufficient to ensure the graphicality of joint degree sequences according to the extended configuration model. For example, consider an ordered joint degree sequence describing ordinary edges and triangles $\{s,t\} =\{(0,1),(1,0),(1,0)\}$ which is graphic according to the EGI, Eq \ref{eq:EGT}, and the HL applied to the overall edges, but is not according to the extended configuration model. We require the EGT to hold among the ordinary edges such that $\sum_is_i=2\mathcal H$ where $\mathcal H\in \mathbb Z$ is the number of ordinary edges and that
\begin{equation}
    \sum_{i=1}^{n} s_i \leq n(n-1)+\sum_{i=n+1}^N {\min}(s_i,n)
\end{equation}
holds for $n\in [1,\mathcal N-1]$. For the triangle degree sequence to be graphical, we require that the sum of the number of triangles is divisible by 3 
\begin{equation}
    \sum_{i=1}^{\mathcal N}t_i=3\mathcal T
\end{equation}
which is a modified handshaking lemma, as well as a modified inequality
\begin{equation}
    2\sum_{i=1}^{n} t_i \leq n(n-1)+\sum_{i=n+1}^N {\min}(2t_i,n)\label{eq:modEGT}
\end{equation}
must hold for $n\in [1,\mathcal N-1]$. The factor of 2 in Eq \ref{eq:modEGT} is due to the characteristic scale of triangles. Together these conditions form an extended Erd\H{o}s-Gallai theorem and they ensure that the joint degree sequence is graphic. The conditions for the graphicality of joint degree sequences of configuration models comprising higher-order cliques can now be written by exploiting the characteristic size of each clique. 


\section{Simple cycles}
\label{sec:weak}

In this section we derive a formula for the probability that a node fails to become attached to the GCC despite its involvement in a simple cycle of length $N$. We define a \textit{complete} graph to indicate that all edges in the cycle are intact; whilst a \textit{connected} graph is one in which there exists at least one pathway between all nodes. The removal of a single edge from a cycle with $N>2$ will ruin the complete property, but it will still be fully connected.

In the following we reserve $j$ for an index over the number of edges we have removed from the cycle and we reserve $r$ for an index over the number of nodes $n$ we have removed from the cycle. 

We begin by defining the probability of the complete cycle. In this case, since all edges are present, each node must not belong to the GCC and hence we have 
\begin{equation}
    u^{N-1}T^{N}
\end{equation}
We can remove an edge from the cycle and still retain full connectivity among the nodes. Given that there are $N$ edges we obtain 
\begin{equation}
    N(uT)^{N-1}(1-T)
\end{equation}
If another edge is removed, then a node can become isolated and hence we must reduce the power of $u$ by one to obtain 
\begin{equation}
    (N-1)(uT)^{N-2}(1-T)^2
\end{equation}
The leading factor of $(N-1)$ accounts for the number of edges to remove. The removal of $j$ edges yields 
\begin{equation}
    \sum_{j=0}^{N-2} (j+1)(uT)^j(1-T)^2
\end{equation}
Therefore, the entire expression for $g_\tau$ for weak cycles is 
\begin{align}
    g_\tau=\ &u^{N-1}T^{N} + N(uT)^{N-1}(1-T)\nonumber\\
    &+ \sum_{j=0}^{N-2} (j+1)(uT)^j(1-T)^2\label{eq:simple}
\end{align}


\section{Cliques}
\label{sec:cliques}

\begin{figure}[ht!]
\centering
\includegraphics[width=0.185\textwidth]{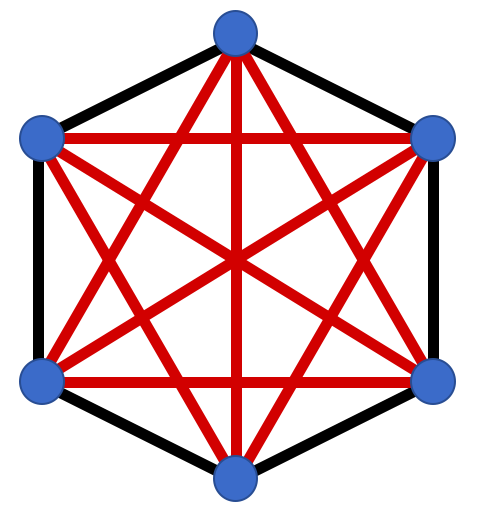}
\caption{The 6-clique has 6 nodes (blue), 6 exterior edges (black) and $6(6-1-2)/2=9$ interior edges (red). There are $6(6-1)/2=15$ edges in total. }
\label{fig:1}
\end{figure}

In this section we derive an exact expression for the probability $g_N$ that a node fails to become attached to the GCC when it is a constituent of a clique of size $N$. Cliques have been studied previously using alternative methods; however, these approaches use recursion to obtain a solution \cite{PhysRevE.68.026121,allard_hebert-dufresne_young_dube_2015}. As with the weak case, we frame our theory in layers around integer powers of $u$ in the range $[0,N-1]$. We categorise the edges of the clique as either \textit{exterior} or \textit{interior} edges, depending on whether they belong to the outer skeleton of the cycle or connect nodes across the interior, through the shape respectively, see Fig \ref{fig:1}.

\begin{figure}[ht!]
\centering
\includegraphics[width=0.185\textwidth]{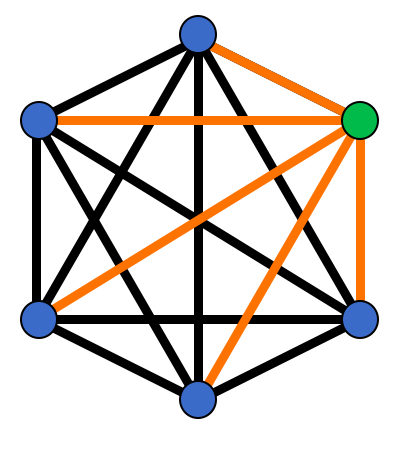}
\caption{The (6-1)-semi-complete clique has 1 coloured node (green) and $6-1=5$ ordinary nodes. The $(6-1)=5$ edges that emanate from the coloured node have also been coloured (orange). If we were to ignore colouring, this cycle would be the 6-clique. 
}
\label{fig:2}
\end{figure}

 We define another term, a \textit{$(N-n)$-semi-complete} graph to be the complete clique of codimension-$(n)$ embedded in the clique of size $N$ with $n$ nodes, and their edges, coloured. In other words, an $(N-1)$-semi-complete clique is a clique of size $N$ with $1$ node coloured, and all edges that connect to the coloured node are also coloured, see Fig \ref{fig:2}. A $(N-2)$-semi-complete clique is a clique of size $N$ with $2$ coloured nodes, whose edges to all other nodes (and between the coloured nodes themselves) are also marked. 

With these definitions in place, let us begin the derivation. The first and arguably the easiest layer is the fully connected graph of size $N$. With all of its edges intact we pick a focal node and set the remaining $(N-1)$ nodes to the $u$ state. The fully connected, complete clique of size $N$ occurs with probability 
\begin{equation}
    u^{N-1}T^NT^{N(N-1-2)/2}\label{eq:complete}
\end{equation}
Examining these terms, we note that all nodes other than the focal node must not be in the GCC if all of their edges are occupied. There are $N$ exterior edges and $N(N-1-2)/2$ interior edges. There is only one way to pick this shape, so its \textit{multiplicity} (the number of different ways the configuration can occur) is unity. 

As remarked above, for $N>2$, we can remove edges from this cycle and it will still be fully connected, although no longer complete. It happens that we can remove all of the interior edges, and even one of the exterior edges and still make connected graphs. If we set one of the interior edges unoccupied, we have 
\begin{equation}
    q_{N,N(N-1)/2-1} u^{N-1}T^N T^{N(N-1-2)/2-1}(1-T)
\end{equation}
where $q_{m,k}$ is the number of connected graphs that can be formed over $m$ labelled nodes with $k$ edges (see Appendix \ref{sec:appendixA}).

If we remove a second edge we have
\begin{equation}
    q_{N,N(N-1)/2-2} u^{N-1}T^N T^{N(N-1-2)/2-2}(1-T)^2
\end{equation}

The removal of $j$ edges is now given by 
\begin{align}
    &\sum^{\mathcal E(N)}_{j=1} q_{N,N(N-1)/2-j} u^{N-1}T^NT^{N(N-1-2)/2-j}\nonumber\\
    &\times(1-T)^j\label{eq:complete_m_j}
\end{align}
where $\mathcal E(N)=N(N-1-2)/2+1$. 

If we were to remove another edge from the graph, we would isolate a node, and this will decrease the largest power of $u$ by one. There are $(N-1)$ nodes that we could remove and all edges that point to the removed node must now be $(1-T)$, of which there are $(N-1)$. Putting this together the $(N-1)$-semi-complete graph, or codimension-1 subgraph in the $N$-clique occurs with probability 
\begin{equation}
    (N-1) u^{N-2} T^{N-2}T^{(N-1)(N-1-1-2)/2}(1-T)^{N-1}\label{eq:cd1scm1}
\end{equation}
where the number of interior edges among the non-removed nodes is now ${(N-1)(N-1-1-2)}/{2}$. We can imagine this as a clique of size $(N-1)$ embedded within the $N$-clique, and the remaining edges are set to $(1-T)$. We recall the $(6-1)$-semi-complete graph from Fig \ref{fig:2}, the removed node is green and the $(1-T)$ edges are orange. The leading factor of $(N-1)$ in Eq \ref{eq:cd1scm1} accounts for the choices of node we could remove other than the focal node.

As with the complete case, we can remove edges from this graph and still retain connectivity among the $(N-1)$ non-removed nodes. Removal of a single edge occurs with probability 
\begin{align}
    &(N-1)q_{N-1,X_{N-1,1}} u^{N-2}T^{N-2}T^{(N-1)(N-1-1-2)/2-1}\nonumber\\
    &\times(1-T)^{N-1+1}
\end{align}
where $X_{N-r,j}$ is the number of edges in the $(N-r)$-clique minus $j$ 
\begin{equation}
    X_{N-r,j} = (N-r)(N-r-1)/2-j
\end{equation}
Let us remove a second edge from this cycle to obtain 
\begin{align}
    (N-1)&q_{N-1,X_{N-1,2}} u^{N-2}T^{N-2}T^{(N-1)(N-4)/2-2}\nonumber\\
    &\times (1-T)^{N-1+2}
\end{align}
The removal of $j$ edges now proceeds as 
\begin{align}
    (N-1)&\sum^{\mathcal E(N-1)}_{j=1} q_{N-1,X_{N-1,j}}u^{N-2}T^{N-2}T^{(N-1)(N-4)/2-j}\nonumber\\
    &\times (1-T)^{N-1+j}\label{eq:cd1sc}
\end{align}
To be clear, this is the equation of the $N$-clique with one node removed and up to $j=(N-1)(N-4)/2+1$ edges removed.

\begin{figure}[ht!]
\centering
\includegraphics[width=0.185\textwidth]{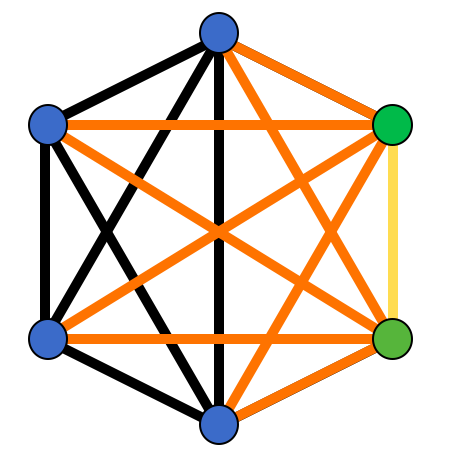}
\caption{The (6-2)-semi-complete clique has 2 coloured nodes (green) and $6-2=5$ ordinary nodes (blue). The $(N-1) + (N-2)=9$ edges that emanate from coloured nodes have also been coloured (orange). Notice that the edge that connects the two coloured nodes (yellow) has been coloured differently than the other edges. Interface edges connect blue nodes to green nodes. There are 9 - 1 interfaces edges in this example.
}
\label{fig:3}
\end{figure}

Further removal of an edge would isolate a node and hence, we claim that this level is now completed. Although the pattern is largely the same as above, there is a complexity with the removal of a second node. Currently, we absorb all of the removed node's edges into the $(1-T)$ box. However, when a second node is removed, there is a connection between the removed nodes that need not be $(1-T)$. Therefore, we must subtract from this power those connections between removed nodes. This is simply the number of edges in a clique of size equal to the number of removed nodes, $n$. We introduce the term \textit{interface edges} to be edges that connect removed nodes to non-removed nodes, see Fig \ref{fig:3}.

The number of interface edges is given by the total number of edges that the removed nodes have, minus the number of edges that connect removed nodes to each other. If there are $n<N$ removed nodes, then there are a total of 
\begin{equation}
    \sum^n_{i=1}(N-i) \nonumber
\end{equation}
coloured edges (orange plus yellow), of which a total of
\begin{equation}
    \frac {n(n-1)}{2}\nonumber
\end{equation}
will point to other removed nodes (yellow). Hence, the total number, $\omega(r)$, of interface edges (orange) is \begin{equation}
    \omega(r) = \sum^r_{i=1}(N-i) - \frac {r(r-1)}{2}
\end{equation}
Hence, for subsequent node removals, the number of edges that are required to be $(1-T)$ are given by the number of interface edges. In the case that $n=1$ we find that the number of interface edges is $N-1$, in agreement with the previous workings. 

We will now remove a second node from the $N$-clique and we begin by describing the $(N-2)$-semi-complete graph, from which we will then remove edges in a sequential and now hopefully familiar fashion. 

There are $(N-1)$ ways to remove the first node followed by $\binom{N}{2}$ ways to remove the second node, so the binomial coefficient will lead the expression. The chain of nodes not in the GCC now occurs with probability $u^{N-3}$ comprising $N$ less two removed nodes and one focal node. The outer $T$ skeleton of the $(N-2)$-semi-completed graph has probability $T^{N-3}$. The number of interior edges among present nodes is then $(N-2)(N-2-1-2)/2$. The number of interface edges is
\begin{equation}
 \sum_{i=1}^2(N-i)-\frac{2(2-1)}{2} = 2(N-2)\nonumber
\end{equation}
All together, the expression for a clique of size $N$ with 2 removed nodes (a semi-complete graph of codimension 2) is given by 
\begin{align}
    &\binom{N-1}{2}u^{N-3}T^{N-3} T^{(N-2)(N-2-1-2)/2}\nonumber\\
    &\times(1-T)^{2(N-2)}\label{eq:cd2sc}
\end{align}
We can then remove all of the interior edges among the non-removed nodes, as well as a single exterior edge and place them into the $(1-T)$ box. Removing one edge we have 
\begin{align}
    &\binom{N-1}{2}q_{N-2,X_{N-2,1}}u^{N-3}T^{N-3} \nonumber\\
    &\times T^{(N-2)(N-2-1-2)/2-1}(1-T)^{2(N-2)+1}
\end{align}
All of the interior edges of the non-removed subgraph can be removed, along with one exterior edge, and still permit connected subgraphs of size $(N-n)$ among the non-removed nodes. Hence, the removal of $j$ such edges yields
\begin{align}
   &\binom{N-1}{2}\sum_{j=1}^{\mathcal E(N-2)}q_{N-2,X_{N-2,j}}u^{N-3}T^{N-3}\nonumber\\
    &\times T^{(N-2)(N-2-1-2)/2-j} (1-T)^{2(N-2)+j}\label{eq:cd2mj}
\end{align}
Subsequent loss of edges will isolate a further node. 

We have now encountered all the sufficient logic that we require for the correct abstraction of the formula to account for arbitrary numbers of removed nodes and edges from a clique of size $N$.

For a clique of size $N$, let there be $n$ removed nodes. There are 
$\binom{N-1}{r} $ ways to remove the $r\leq n$ nodes sequentially. The power of $u$ is given by $(N-r-1)$; this is the power of the exterior $T$ also; the interior power of $T$ is given by $\mathcal (N-r)$. The final expression therefore is given by 
\begin{align}
g_N =\ & \sum_{r=0}^{N-1}\binom{N-1}{r} \sum^{\mathcal E (N-r)}_{j=0}q_{N-r,X_{N-r,j}}(uT)^{N-r-1} \nonumber\\
&\times T^{\mathcal E (N-r)-1-j}(1-T)^{\omega(r)}
\label{eq:mainMainFINAL}
\end{align}
This equation is the main result of this section. 

\subsection{Percolation threshold}
\label{subsec:percthresh}

We now turn our attention to the location of the critical point for the formation of a GCC among networks comprised entirely of $N$-cliques during bond percolation. From section \ref{sec:introduction} we understand that in order to obtain the percolation properties of the network, we have to evaluate the derivative of $g_\tau$ with respect to $u$. This derivative is found to be 
\begin{align}
\frac{\partial g_N}{\partial u} =\ & \sum_{r=0}^{N-1}\binom{N-1}{r}(N-r-1) \sum^{\mathcal E (N-r)}_{j=0}q_{N-r,X_{N-r,j}} \nonumber\\
&\times(uT)^{N-r-2} T^{\mathcal E (N-r)-j}(1-T)^{\omega(r)}
\label{eq:derivativeCLIQUE}
\end{align}
The percolation threshold is then obtained when $u=1$, and following a similar analysis to the tree-like topology we obtain 
\begin{equation}
    \frac{\partial g_N}{\partial u}\bigg|_{u=1} \frac{\langle k^2-k\rangle}{\langle k\rangle} =1\label{eq:MRC}
\end{equation}

For example, the derivative for 3-cliques is found to be 
\begin{equation}
    \frac{\partial g_3}{\partial u} = 2T(1-T)^2+6uT^2(1-T)+2uT^3
\end{equation}
Evaluated at $u=1$ and inserted into Eq \ref{eq:MRC} we have $2( T^2 + T - T^3)\langle t\rangle - 1 = 0$ where $\langle t \rangle $ is the average number of triangles that a node belongs to; and, we have assumed that the cycles are Poisson distributed. Using Gauss's lemma, this cubic expression is reducible in $T$ into the quadratic form whose roots yield the critical transmissibilities of the model, and hence, the critical point occurs at
\begin{equation}
    T^* = -1+ \frac{1}{2}\sqrt{4  + \frac{4}{\langle t\rangle}}
\end{equation}
We repeat the calculation for the 4-clique to obtain the following polynomial
\begin{equation}
    \frac{\partial g_4}{\partial u}=3T(-2T^5+7T^4-7T^3+2T+1)
\end{equation}
The Galois group of the quintic part is the symmetric group, $S_5$, and hence a root cannot be found. It is unlikely that percolation properties of higher-order cycles can be resolved analytically due to the  Abel-Ruffini theorem. 

In conclusion, we have derived an exact formula to obtain the bond percolation properties, including the size of the GCC and the location of the percolation threshold, of configuration model networks comprised of higher-order subgraphs. We presented our method for simple cycles and cliques, however, a wide range of subgraphs can also be considered. We have also studied the conditions for degree sequences to be considered graphic for these networks and found the correct extension of the Erd\H{o}s-Gallai theorem.

\section{ACKNOWLEDGMENTS}

The authors would like to thank an anonymous referee from an earlier paper for providing Eq \ref{eq:simple} to us and thus igniting the pathway to the exact formula. We would also like to thank the School of Chemistry and the School of Biology of the University of St Andrews for funding this work.


\bibliography{bib}

\appendix
\section{$q_{n,k}$}
\label{sec:appendixA}

The number of connected graphs of $N$ labelled vertices over $k$ edges is given by $q_{n,k}$. This quantity has a well known recursion formula as well as a closed-form analytical solution \cite{wilf_1994,riddell_uhlenbeck_1953,PhysRevE.68.026121}. Given the importance of this quantity to the contents of this paper, we will review this derivation now. 

Let $\mathcal{Q}$ be the combinatorial class of connected graphs and $\mathcal{G}$ the combinatorial class of all labelled graphs. The relation between these two classes is the set-of relation: a graph is a set of connected components. This indicates that the mixed exponential generating function $G(z)$ of $\mathcal{G}$ can be generated from $Q(z)$ according to the following relationship 
\begin{equation}
    G(z) = \exp Q(z)
\end{equation}
We can readily compute $G(z)$ as 
\begin{align}
G(z) =\ & 1 + \sum_{m\ge 1} (1+u)^{m(m-1)/2} \frac{z^m}{m!}
\end{align}
This yields an expression for the entire series of connected graphs, $Q(z)$, since, $Q(z)= \log G(z)$ such that we obtain
\begin{align}Q(z) = \sum_{q\ge 1} (-1)^{q+1} \frac{1}{q}
\left(\sum_{m\ge 1} (1+u)^{m(m-1)/2} \frac{z^m}{m!}\right)^q
\end{align}
We now examine the case of $n$ nodes and $k$ edges where $k\ge n-1$ by extracting the coefficient $q_{n,k}$ of $[z^n] [u^k]$. Note that the term in the parenthesis has minimum degree $q$ in $z$, allowing us to disregard the series beyond $q>n.$ This yields the formula for the number of connected labelled graphs with $n$ nodes and $k$ edges as
\begin{align}
    q_{n,k} =\ & n! [z^n] [u^k] \sum_{q=1}^n (-1)^{q+1} \frac{1}{q}
\nonumber\\
& \times \left(\sum_{m=1}^n (1+u)^{m(m-1)/2} \frac{z^m}{m!}\right)^q
\end{align}
As an example of $q_{n,k}$ in Eq \ref{eq:mainMainFINAL}, we examine the coefficients of the 4-clique when there are no removed nodes, that is, when $n=0$. 
\begin{figure}[htbp!]
\centering
\includegraphics[width=0.3\textwidth]{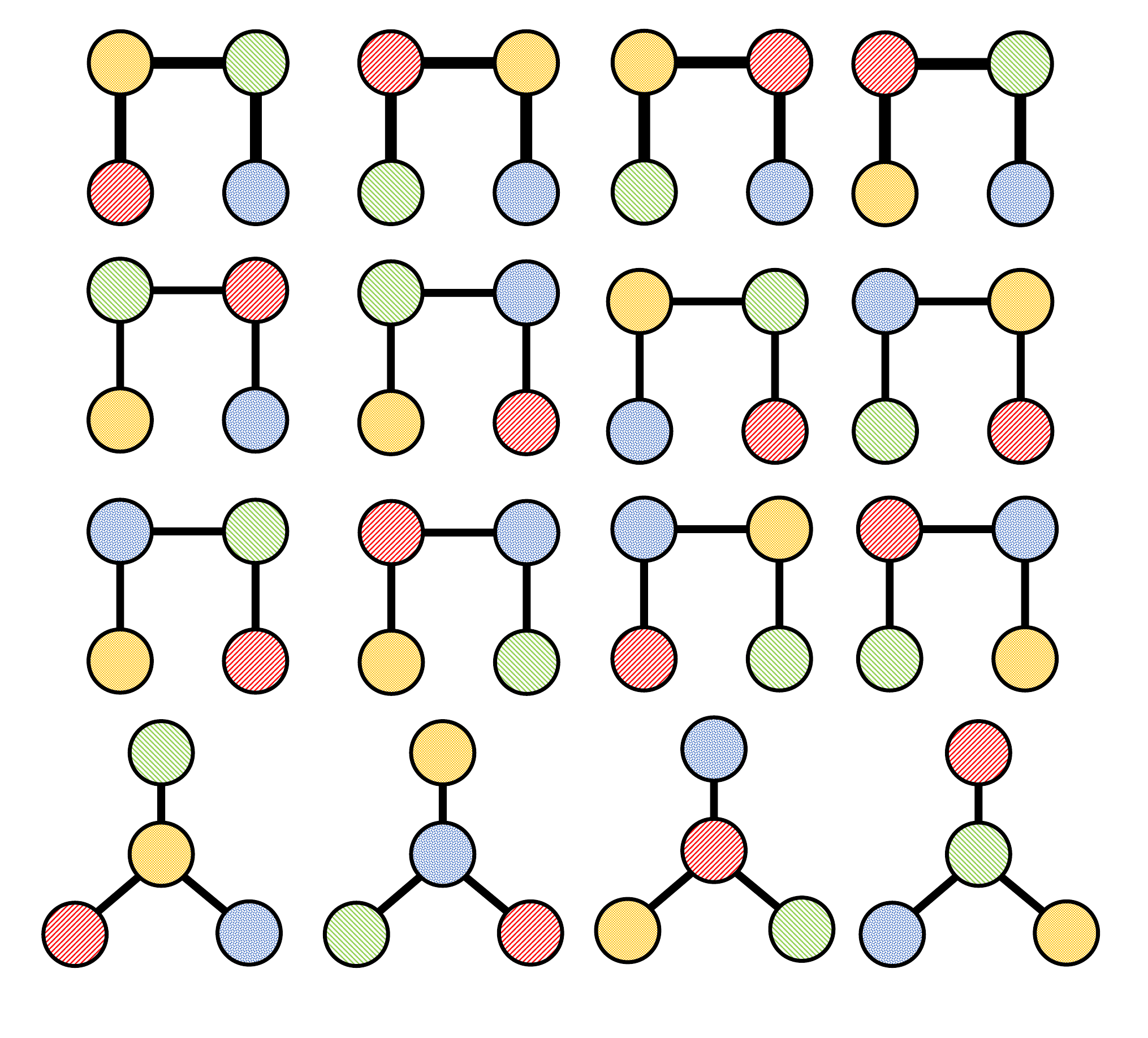}
\caption{The 16 graphs that can be made among $4$ labelled nodes with $3$ edges is given by $q_{4,3}$.  
}
\label{fig:5}
\end{figure}
From table \ref{tab:1}, we observe the leading coefficients of the terms in $u^3$ are $q_{4,k}=1,6,15$ and $16$ which correspond to the number of graphs that can be made with $k=6,5,4,3$ edges, respectively. The set of graphs that can be made from $q_{4,3}$ is presented in Fig \ref{fig:5}.

\begin{widetext}
\vspace{10cm}
\pagebreak
\section{Displayed Clique formulas}

The expressions for cliques of 6 nodes or fewer are shown in table \ref{tab:1}. It is clear upon comparison that the $q_{n,k}u^xT^y$ structure of the polynomials appear to repeat across the orders of increasing clique size. For instance, the polynomial of $u^4$ can be compared between the $N=5$ and $N=6$ equations; however, in each case, the exponent of the interface edges, $(1-T)^z$, varies. We further note that Eqs \ref{eq:simple} and \ref{eq:mainMainFINAL} are in agreement for $N=3$.

\begin{center}
\begin{table}
\centering
\renewcommand{\arraystretch}{1.2}
\setlength{\tabcolsep}{1.3em}
\begin{tabular}{ c||c }  
$N$-Clique & $g_N$ equation \\  [1.3ex]
\toprule  
 \\ 
 3 & $(1-T)^2+2uT(1-T)^2 + 3(uT)^2(1-T)+u^2T^3$ \\  [1.3ex]
  \hline\\ 
4 & 
$\begin{aligned}[t] 
&(1-T)^3 + 3uT(1-T)^4 +3u^2(T^3(1-T)^3+3T^2(1-T)^4)\nonumber\\
&+ u^3(T^6+6T^5(1-T) + 15T^4(1-T)^2+16T^3(1-T)^3)
\end{aligned}$\\
\\ 
  \hline  \\ 
 5 & $\begin{aligned}[t] 
&(1-T)^4 + 4uT(1-T)^6 + 6u^2(T^3(1-T)^6+3T^2(1-T)^7)+4u^3(T^6(1-T)^4 + 6T^5(1-T)^5\nonumber\\
&+15T^4(1-T)^6 + 16T^3(1-T)^7) +u^4(T^{10} + 10T^9(1-T) + 45T^8(1-T)^2+120T^7(1-T)^3\nonumber\\
&+ 205T^6(1-T)^4 + 222T^5(1-T)^5 + 125T^4(1-T)^6)
\end{aligned}$\\
\\ 
  \hline \\ 
  6 &
$\begin{aligned} 
& (1-T)^5 +  5uT(1-T)^8 + 10u^2(T^3(1-T)^9 + 3T^2(1-T)^{10}) + 10u^3(T^6(1-T)^8+ 6T^5(1-T)^9\nonumber\\
& + 15T^4(1-T)^{10}+16T^3(1-T)^{11}) + 5u^4(T^{10}(1-T)^5+10T^9(1-T)^6+45T^8(1-T)^7\nonumber\\
&+120T^7(1-T)^8
+205T^6(1-T)^9+222T^5(1-T)^{10}+125T^4(1-T)^{11}) + u^5(T^{15} + 15T^{14}(1-T)\nonumber\\
& + 105T^{13}(1-T)^2 + 455T^{12}(1-T)^3+1365T^{11}(1-T)^4+2997T^{10}(1-T)^5 + 4945T^9(1-T)^6\nonumber\\
& + 6165T^8(1-T)^7 +5700 T^7(1-T)^8+3660T^6(1-T)^9+1296T^5(1-T)^{10}
\end{aligned}$

\end{tabular}
\centering
\caption{The $g_N$ expressions for cliques of 6 vertices or fewer obtained from Eq \ref{eq:mainMainFINAL}.}
\label{tab:1}
\end{table}
\end{center}
\end{widetext}

\end{document}